\documentclass[useAMS,usenatbib]{mn2e}
\usepackage{graphicx}
\usepackage{rotating,times,pictex,graphicx,latexsym}
\usepackage{color}
\usepackage{longtable,amsmath}
\usepackage{lscape}

\title{FSR\,1735 -- A new globular cluster candidate in the inner Galaxy} 

\author[Froebrich, Meusinger \& Scholz]{D.~Froebrich$^{1}$\thanks{Based on
observations collected at ESO, Chile; ESO  077.B-0074(A)}\thanks{E-mail:
df@star.kent.ac.uk}, H.~Meusinger$^{2}$ and A.~Scholz$^{3,4}$\\ $^1$ Centre for
Astrophysics and Planetary Science, University of Kent, Canterbury, CT2 7NH, UK
\\ $^2$ Th\"uringer Landessternwarte Tautenburg, Sternwarte 5, 07778 Tautenburg,
Germany \\ $^3$ University of St. Andrews, Department of Physics \& Astronomy,
North Haugh, St. Andrews, Fife, KY16 8SS, Scotland, UK \\ $^4$ University of
Toronto, Department Astronomy \& Astrophysics, 50 St. George Street, Toronto,
Ontario M5S 3H8, Canada} 

\begin{document}

\date{Received sooner; accepted later}
\pagerange{\pageref{firstpage}--\pageref{lastpage}} \pubyear{2007}
\maketitle

\label{firstpage}

\begin{abstract}

We carry out a large program to classify newly discovered star clusters from
Froebrich et al. \cite{2006MNRAS.subm.F} in the inner Galaxy. Here, we  present
a first analysis of a new high-priority globular cluster candidate, FSR\,1735 at
$l$=339.1879; $b=-1.8534$, based on new deep near infrared observations from
Sofi at the NTT. A significant peak in the K-band luminosity function is found,
which is interpreted as the clump of post-He-flash stars. The distance and the
reddening of the cluster are determined to 9.1\,kpc and $A_K$=0.5\,mag,
respectively, the metallicity is estimated to [M/H]=$-0.8$. Radial star density
profiles are used to measure the core radius and the tidal radius of the
cluster. 
The lack of signs for on-going star formation and the position in the Galaxy
pose strong arguments against the interpretation of this object as a young or
old open cluster. All the observational evidence is in agreement with the
interpretation that FSR\,1735 is a so far unknown globular cluster in the inner
Galaxy. 

\end{abstract}


\section{Introduction}

Star clusters provide us with unique laboratory conditions to investigate
various aspects of astrophysics. They represent groups of stars with similar
ages, metallicities and distances. Beside testing stellar evolutionary models
they also enable us to trace the distribution of interstellar material along the
line of sight. Orbits of star clusters around the Galactic centre can be used to
probe the distribution of mass in the Galaxy. Galactic open clusters (OCs) 
trace the recent star formation history in the Galaxy, while globular clusters 
(GCs) can be used to gather information about the formation and early  evolution
of the Milky Way. The differences in the formation of these  cluster types is
reflected in the distribution on the sky. While  OCs are mainly distributed in
the Galactic plane, GCs are  concentrated towards the Galactic centre. 

The number of known OCs has risen dramatically in recent years with the
availability of deep large scale near infrared surveys such as 2MASS (Skrutzkie
et al. \cite{2006AJ....131.1163S}). Numerous surveys in the 2MASS catalogue were
conducted to identify new star clusters  (e.g. Carpenter et
al.\cite{2000ApJS..130..381C};  Dutra \& Bica
\cite{2000A&A...359L...9D},\cite{2001A&A...376..434D};  Ivanov et al.
\cite{2002A&A...394L...1I};  Bica et al. \cite{2003A&A...404..223B}; Kronberger
et al. \cite{2006A&A...447..921K}; Froebrich et al. \cite{2006MNRAS.subm.F}).
Several hundred new clusters and cluster candidates were identified by these
works,  preferably close to the Galactic plane. 

Contrary to OCs, the total number of known Milky Way\,GCs has increased by just
a few percent over the last 25 years. So far 152 Milky Way\,GCs are known, but
this sample is very likely not complete. The "Catalog of Parameters for Globular
Clusters in the Milky Way" by Harris \cite{1996AJ....112.1487H} lists 147
objects. In recent years, this sample was enlarged by five (perhaps six)
objects: 2MASS\,GC01/02 (Hurt et al. \cite{2000AJ....120.1876H}); ESO\,280-SC06
(Ortolani et al. \cite{2000A&A...361L..57O}); GLIMPSE-C01 (Kobulnicky et al.
\cite{2005AJ....129..239K}); GC\,Whiting1 (Carraro et al.
\cite{2005ApJ...621L..61C}); and perhaps SDSS\,J1049+5103 (Willman et al.
\cite{2005AJ....129.2692W}). Originally most of the Milky Way\,GCs were
discovered by visual inspection of photographic plates obtained at optical
wavelengths. Close to the Galactic plane, however, many objects remained
undiscovered due to obscuration from dust. This area, well known as the 'Zone of
Avoidance' in extragalactic astronomy, might have a number of so far unknown
Milky Way\,GCs in store. Based on the spatial distribution of the known GCs
Ivanov et al. \cite{2005A&A...442..195I} estimate a lower limit of $10 \pm 3$
unknown GCs near the Galactic plane ($|z| \le 0.5$\,kpc) and
within 3\,kpc from the Galactic centre. The Zone of Avoidance is obviously
expected to be the least complete area in the Milky Way\,GC sample, though the
recent serendipitous discoveries of the two off-plane clusters GC\,Whiting1
(Carraro et al. \cite{2005ApJ...621L..61C}) and ESO\,280-SC06 (Ortolani et al.
\cite{2000A&A...361L..57O}) have demonstrated that the previous sample of
off-plane clusters was also incomplete.

We are currently conducting a survey of star cluster candidates (selected from
Froebrich et al. \cite{2006MNRAS.subm.F}) to identify a substantial fraction of
the missing Milky Way\,GCs in the Zone of Avoidance and to determine their
properties such as distance, size, extinction, metallicity and age, based on
deep NIR photometry. Here we present the observations of a first very promising
new GC candidate uncovered by this survey. The cluster has the number 1735 in
the list of Froebrich et al. \cite{2006MNRAS.subm.F} and we will hence refer to
the object as FSR\,1735. Its accurate position is $l$=339.1879; $b$=-1.8534 or
$\alpha = 16^{\rm h}52^{\rm m}10.6^{\rm s}, \delta = -47^{\circ}03'29''$
(J2000), about half an arcminute south of the position given in Froebrich et al.
\cite{2006MNRAS.subm.F}.

The paper is structured as follows. In Sect.\,\ref{data} we present our data,
followed by the determination of the properties of the star cluster in
Sect.\,\ref{position}. This is followed by a discussion of the classification of
the cluster in Sect.\,\ref{classification} and the conclusions.

\section{Data}

\label{data}

As part of a larger program to study selected cluster candidates from Froebrich
et al. \cite{2006MNRAS.subm.F} in the Zone of Avoidance, observations with Sofi
at the NTT were carried out on the 7th of May in 2006. We observed in Large
Field mode with a pixel size of 0.288" and a field of view of 4.9'x4.9'. In each
of the three broad band filters (JHK$_s$) a total per pixel integration time of
225\,s was achieved with short 5x5\,s exposures. Standard procedures were used
for flatfielding, sky-subtraction and mosaicing. 

The seeing in the co-added mosaics is 0.85". Source detection and photometry in
the images was done using the SExtractor software (Bertin \& Arnouts
\cite{1996A&AS..117..393B}). Only objects that are detected above the three
sigma noise level in all three filters are used in the subsequent analysis.
Conditions were photometric and photometric calibration of the images was done
using the variety of about 500 2MASS sources in the field around the cluster.
The completeness limits (determined as the peak in the luminosity function)
outside the cluster area are 18.1, 17.1, and 16.7\,mag for JHK, respectively.
Near the cluster centre we are much more hampered by source confusion and the
completeness limits are 17.2, 16.3, 15.7\,mag for JHK, respectively. Typical
photometric errors for the latter brightnesses are 0.045, 0.047, 0.041\,mag for
JHK, respectively.

\begin{figure}
\centering
\includegraphics[width=6.5cm]{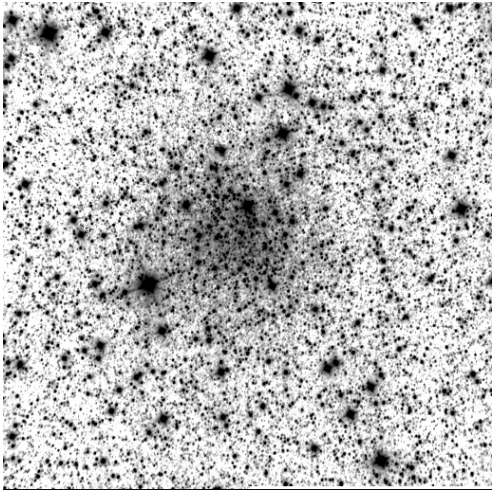}
\caption{\label{kgray} Logarithmic gray scale image of the cluster FSR\,1735.
The image size is 4.9'x4.9'. North is to the top and East to the left.} 
\end{figure}

\section{Results}

\label{position}

\subsection{Basic properties}

Our new deep JHK images show that the nebulous appearance of the cluster in
2MASS images is due to a large number of faint unresolved stars. In
Fig.\,\ref{kgray} we show the logarithmic gray-scale version of the K-band image
of the cluster. The large number of stars in the cluster compared to the
surrounding field can immediately be seen, as well as its almost circular
appearance.

\begin{figure}
\centering
\includegraphics[height=7.5cm,angle=-90]{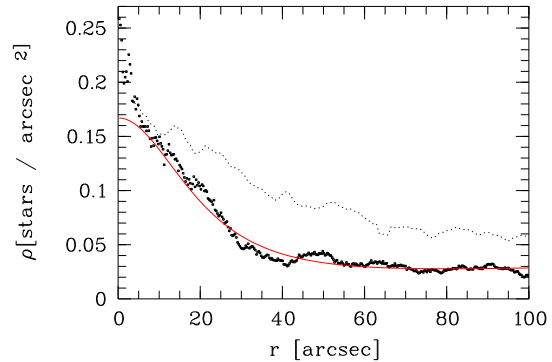}
\caption{\label{dens} Observed K-band star density profile (dots) and best
fitting King-profile (solid line) of the cluster. Overplotted as dotted line is
the star density profile of the unresolved cluster population in arbitrary
units.} 
\end{figure}

We determined the radial density profile of stars one magnitude above the local
completeness limit in the cluster area (i.e. at $m_K=14.7$\,mag). This ensures
that we are not too hampered by a possibly lower completeness limit in the very
centre of the cluster. The resulting star density profile in Fig.\,\ref{dens}
shows that a high number of bright stars is concentrated towards the centre of
the cluster (note that the measured profile is smoothed with a 4.3" radius). The
star density follows a King profile, with the best fit parameters:
$R_{core}=23.5$", $R_{tid}=86$", $I_{cen}=0.27$**/arcsec$^2$,
$I_{back}=0.0278$**/arcsec$^2$. Integrating the King profile till the tidal
radius results in about 300 stars brighter than $m_K$=14.7\,mag in the cluster.
Only inside 4" from the centre the star density exceeds the King profile due to
three bright stars around the central position. 

We determined the radial profile of the average flux per pixel of the unresolved
cluster population of stars by masking all detected stars. This profile is
overplotted in arbitrary units as dotted line in Fig.\,\ref{dens}. One can
interpret this as the number of unresolved stars per pixel, i.e. the density of
faint stars in the cluster. Hence, Fig.\,\ref{dens} allows us to compare the
star density profile of the bright and faint stars  in the cluster. From a King
profile fit we determine $R_{core}=41.5"$ and $R_{tid}=125"$ for the unresolved
stars. The apparent difference cannot be interpreted in terms of mass
segregation due to the small difference in mass between giant - subgiant stars
and (upper) main sequence stars. It has to be attributed to magnitude migration
from faint to bright stars due to the significant crowding (larger than 30\,\%
in the inner 10") even at those bright magnitudes.

\subsection{Reddening}

We constructed colour-colour diagrams of the investigated area. Using only stars
more than 100" away from the cluster centre (control field), we clearly 
identify two groups of objects: 1) stars with $H-K$=0.27\,mag, $J-H$=0.67\,mag,
and $J-K$=1.00\,mag, which represent almost unreddened foreground stars, mostly
K- and M-dwarfs. 2) Objects with redder colours, which represent mostly reddened
background giant stars. The gap in the colour-colour diagram between the two
groups seems to indicate that one single cloud is responsible for most of the
reddening in the field. The MSX Band\,A (6-11\,$\mu$m) image indeed shows a
faint cloud extending across our field of view. We note that no point source is
detected in the MSX images in the area of the cluster.

To determine the extinction along the line of sight to the cluster, we selected
only stars closer than 43" from the cluster centre. In this case only a few
stars show low reddening values, i.e. they are foreground objects. The majority
of objects shows similar colours as the giant stars in the control field. We
determine the mean colour excess of those stars compared to the control field
as: $\left< H-K \right>$=0.24\,mag, $\left< J-H \right>$=0.58\,mag, $\left< J-K
\right>$=0.74\,mag. Using a standard powerlaw for the extinction $A_\lambda
\propto \lambda^{-\beta}$ with $\beta$=1.8 (Draine \cite{2003ARA&A..41..241D}),
we can directly compute the reddening in the K-band. We obtain $A_K$=0.38\,mag,
0.55\,mag, and 0.44\,mag for the $\left< H-K \right>$, $\left< J-K \right>$, and
$\left< J-H \right>$ colour excess values, respectively. Note that other
extinction laws (i.e. $\beta = 1.6$ or using Fitzpatrick
\cite{1999PASP..111...63F}) result in slightly larger values for $A_K$. Hence in
the following we will use a reddening for the cluster of $A_K$=0.5$\pm$0.1\,mag.
Note that the uncertainty in $A_K$ includes any possible reddening of the field
stars whose colours are basically coincident with an unreddened population
(Bessel \& Brett \cite{1988PASP..100.1134B}).

 %
 %

\begin{figure}
\centering
\includegraphics[height=7.5cm, angle=-90]{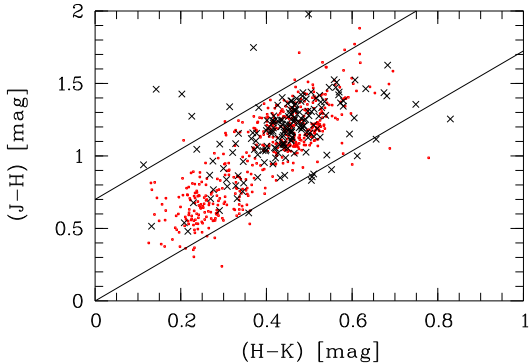}
\caption{\label{colcol} Colour-colour diagram of stars closer than 43" (crosses)
and a control field of stars with distances of more than 100" to the cluster
centre (dots). The solid lines represent the reddening path, i.e. they envelope
the region in which main sequence and giant stars are situated.}
\end{figure}

\subsection{Distance}
 
To determine the distance to the cluster we need to identify objects of known
absolute magnitudes. Due to the metallicity of [M/H]=$-0.8\pm$0.1
(Sect.\,\ref{metall}) the locus of stars in the post-He-flash phase of
evolution, i.e. the horizontal branch (HB) is expected to be degenerated  to a
clump of stars close to the red giant branch (RGB). Additionally, one can use
the stars in the RGB bump, which are pre-He-flash objects. For the absolute
magnitudes of these objects, considering our metallicities, we find $M_K^{\rm
RBG\, bump}=-1.5\pm0.1$\,mag from Valenti et al. \cite{2004A&A...419..139V} for
old globular clusters. For the HB stars we find $M_K^{\rm HB}=-1.4\pm0.1$ for a
6\,Gyr and $-1.2\pm0.1$\,mag for a 12\,Gyr old population (Salaris \& Girardi
\cite{2002MNRAS.337..332S}). Hence, the signals produced by these two features
in the luminosity  function are not widely separated where the HB is expected to
provide a stronger  signal (see e.g. the infrared colour magnitude diagrams of
old clusters  given by Ferraro et al. \cite{2000AJ...119...1282F}).

\begin{figure}
\centering
\includegraphics[height=7.5cm, angle=-90]{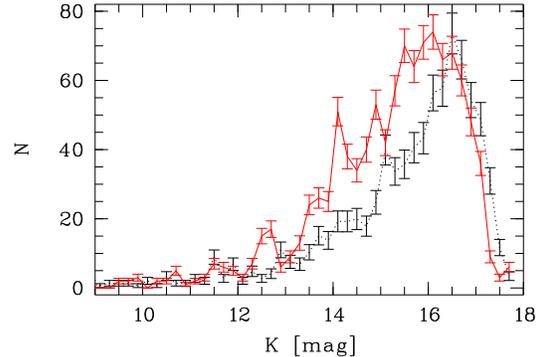}
\caption{\label{khist} Histogram of the apparent K-band magnitude of stars
closer than 43" to the cluster centre (solid line) and outside a radius of 100"
(doted line), normalised to the same area. }
\end{figure}

Figure\,\ref{khist} shows a histogram of the apparent K-band magnitudes of all
stars less than 43" from the cluster centre, and in a colour range $0.3 < (H-K)
< 0.7$. We can compare this to the histogram of objects outside a 100" radius
(normalised to the cluster area). In Fig.\,\ref{khist} one can see that above
$m_K$=12\,mag there are more stars in the cluster than in the control area. Due
to the lower completeness limit in the cluster, above $m_K$=16.5\,mag the number
of stars in the cluster drops below the number in the control field. The
histogram of the cluster stars shows a significant peak at $m_K$=14.1\,mag which
we interpret as the HB stars. We note that this peak appears in the histogram
independent of the binning used. The peak at about $m_K$=15\,mag disappears when
using different bin sizes. There is another peak at $m_K$=12.6\,mag, which is
also significant. It is 0.4\,mag  wide and about 1.5\,mag brighter than the HB.
Alves et al. \cite{2002ApJ...573L..51A} found a similar bump for stars in the
Large Magellanic Cloud and interpreted it as the AGB bump of stars that are at
the onset of helium shell burning. The brightness difference of 5.5\,mag between
the tip of the RGB and the $m_K$=14.1\,mag peak further supports our
interpretation that the latter represents the HB.

Given the absolute magnitudes for the HB stars, we can estimate an extinction
corrected distance modulus of $m-M$=15.0$\pm$0.2 or 14.8$\pm$0.2\,mag for a 6
and 12\,Gyr old population, respectively. This corresponds to a distance from
the Sun of 10.0$\pm$1.0 or 9.1$\pm$1.0\,kpc. With the Galactocentric distance of
8.5\,kpc for the Sun, the Galactocentric distance of the cluster is $R_{GC} =
3.6\pm0.6$ or $3.2\pm0.6$\,kpc. The distance to the Galactic plane is $z=-320$
or $-290$\,pc, and the cluster core diameter is 2.3 or 2.1\,pc.

\subsection{Metallicity}

\label{metall}

\begin{figure}
\centering
\includegraphics[height=8.5cm, angle=-90]{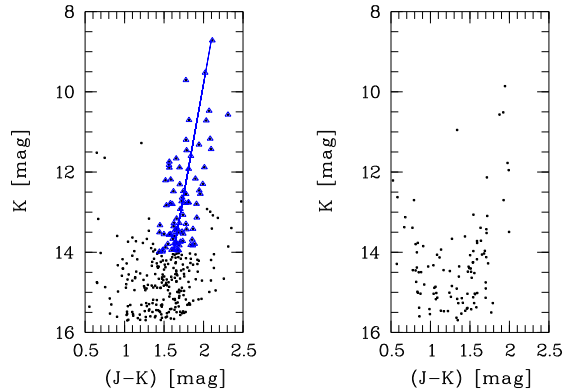}
\caption{\label{colmag} (J-K) vs. K colour-magnitude diagram of stars closer
than 43" from the cluster centre (left) and a control field of stars between
100" and 110" away from the cluster centre (right). The triangles in the left
panel mark all stars brighter than the HB that were used to fit the slope of the
RGB. The best fit is shown as a solid line.}
\end{figure}

One can follow Valenti et al. \cite{2004A&A...419..139V} to determine the
metallicity [M/H] of the cluster. This requires to measure the slope of the RGB
stars in the K vs. (J-K) diagram. Only stars brighter than the HB/RGB bump
should be used. In Fig.\,\ref{colmag} we show the (J-K) vs. K colour-magnitude
diagram. Stars between 100" and 110" away from the cluster centre (control
field) are plotted in the right panel. In the left panel squares indicate all
stars within 43" from the cluster centre. We mark with triangles the stars that
were used to obtain the best fit (solid line). Depending on which stars exactly
we choose to determine the fit, we obtain a range from $-0.09...-0.08$ for the
slope in the diagram. According to Fig.\,8 in Valenti et al.
\cite{2004A&A...419..139V}, this converts into the metallicity range
$-0.7<$[M/H]$<-0.9$. 

\subsection{Integrated Brightness}

To determine the brightness of the entire cluster, we can use two approaches. a)
Integrate the flux from all stars in the area of the cluster, and correct
statistically for foreground/background stars using the histogram in
Fig.\,\ref{khist}. b) Integrate all the flux in the cluster area and correct by
the average flux outside the cluster. Method a) hence determines the integral
brightness of all detected stars, while b) determines the integral light of all
stars in the cluster. The integrated apparent K-band brightness (corrected for
$A_K$) of all detected stars in the cluster is $m^{stars}_K$=8.1\,mag. When
determining the integral dereddened brightness of the entire flux we obtain 
$m^{all}_K$=6.6\,mag. This shows that only 25\,\% of the flux from the cluster
is in the detected stars. The rest is in the unresolved cluster population. If
all the undetected stars were as bright as the extinction corrected completeness
limit in the cluster (i.e. $m_K$=15.2) then we are missing about 2800 stars.
Together with the estimated 300 members brighter than $m_K$=14.7\,mag (see
Sect.\,\ref{position}) the cluster has a minimum number of 3100 member stars.
This is, however, a weak lower limit, since it does not account for the stars
between $m_K$=14.7 and 15.7 and the fact that the unresolved population is
certainly dominated by stars fainter than $m_K$=15.7\,mag.  An independent
estimate of the total number of stars will be discussed in Sect.\,4.3.

With the distance modulus we can convert the apparent dereddened K-band
magnitude of the integrated light ($m^{all}_K$=6.6\,mag) into an absolute
magnitude of $M_K=-8.2\pm$0.2\,mag (assuming a 12\,Gyr old population). This can
be converted into $M_V=-6.3$\,mag using $V-K$=1.9$\pm$0.1\,mag (Leitherer et al.
\cite{1999ApJS..123....3L}).

\section{Classification}

\label{classification}

 %
 %

In the following we will discuss the three possibilities for the classification
of FSR\,1735: 1) young open cluster, 2) old open cluster, 3) globular cluster. 

\subsection{Young open cluster}

To classify Cl\,1715 as a young stellar cluster it should show signs of on-going
or recent star formation. It is at the same line of sight as a faint cloud of
dust seen in emission in the MSX Band\,A image. However, the cloud is much more
extended than the cluster, and the entire area is covered with faint emission
patches in the MSX Band\,A image. Furthermore, there is no bright emission at
far infrared wavelengths visible  at the cluster position in the IRAS images.
This indicates that the dust  cloud is foreground to the cluster, and indeed to
most of the other field  stars. This fact is also supported by the apparent gap
in the colour-colour  diagram (see Fig.\,\ref{colcol}). Hence, there is no
definite indication  that the dust cloud is physically connected to the
cluster. 

There are no MSX or IRAS point sources in the cluster indicating on-going star
formation. In the colour-colour diagram in Fig.\,\ref{colcol} there are 11
objects below the reddening path, which might indicate that these are young
stellar objects. This number is higher than what is expected from the control
field. About 1-2 stars per cluster area in the control field are below the
reddening path. We have hence checked all 11 YSO candidates by eye in order to
verify their colour. All but one object turn out to be close double stars and/or
very faint with problems in the photometry. The reason might be that the
SExtractor software has problems with faint stars in crowded regions of variable
background brightness (this effect accounts as well for the objects above the
reddening path). Hence, only one object remains below the reddening path. Note
that a slight change in the assumed dust properties of the reddening material
can bring almost all of the YSO candidates into the reddening path as well.

There is also no radio continuum source connected to the cluster, which could be
indicating on-going star formation. The closest radio source in the 4.85\,GHz
survey of Wright et al. \cite{1994ApJS...91..111W} is a 48\,$\pm$9\,mJy
detection 4.35' from the cluster centre. 

The cluster is situated close to the Galactic Plane. However, at the computed
distance of 9.1\,kpc it would be 290\,pc below the plane, which is about 5.5
times the scale height of 53\,$\pm$\,5\,pc for open clusters (Joshi et al.
\cite{2005MNRAS.362.1259J}). Hence, the  position of the cluster in the inner
Galaxy in combination with its sub-solar metallicity of [M/H]$=-0.8$ is also a
strong argument against the interpretation of FSR\,1735 as a young open cluster.

\subsection{Old open cluster}

If there are no signs of on-going star formation the object could be an old open
cluster. However, the position in the Galaxy poses a strong argument against
this interpretation. As shown a 6\,Gyr old population would have a
Galactocentric distance of 3.6\,kpc and is situated 320\,pc below the Galactic
plane. While the distance to the plane is in agreement with the scale-height of
375\,pc for old open clusters determined by Janes \& Phelps
\cite{1994AJ....108.1773J}, the distance to the Galactic centre is too low to
interpret FSR\,1735 as an old open cluster. According to Friel
\cite{1995ARA&A..33..381F}, old Galactic clusters are totally absent inside
7.5\,kpc from the Galactic centre. This is caused by frequent encounters with
giant molecular clouds in the inner Galaxy that destroy such objects. Note, that
even if the peak at $m_K = 12.6$\,mag is the HB, the Galactocentric distance of
the cluster would be only 4.2\,kpc.

If it is an old open cluster outside the 7.5\,kpc radius, a distance of about
15\,kpc would be required. Given our reddening and [M/H] values, this is only
possible for an at maximum 0.5\,Gyr old population (Salaris \& Girardi
\cite{2002MNRAS.337..332S}), i.e. $M_K^{HB}=-2.3$\,mag. It is, however,
questionable how such a low metallicity cluster can have formed only 500\,Myrs
ago at this Galactocentric distance. One furthermore expects a number of
luminous supergiants $M_K<-6$ for such a young population. Only one such object
is found. It is thus very improbable that the object is an old open cluster.

\subsection{Globular cluster}
 
Does FSR\,1735 resemble a typical GC? According to Salaris \& Girardi 
\cite{2002MNRAS.337..332S} a 12\,Gyr old population has 4.7$\cdot$10$^{-4}$ HB
stars per solar mass of originally formed stars.  From the star counts we
estimate about 30 HB stars in FSR\,1735 which yields a cluster mass estimate of
6.5$\cdot$10$^4$\,M$_\odot$. Assuming an average mass of 0.5\,M$_\odot$ per 
star, this converts into $\sim$10$^5$ stars in the cluster.  Hence, every
determined cluster parameter (summarised in Table\,\ref{properties}) as well as
the objects position in the Galaxy supports the interpretation of FSR\,1735 as a
so far unknown Milky Way globular cluster. 

All parameter values are well within the typical range of the known GCs in the
Galaxy (e.g. Harris \cite{1996AJ....112.1487H}). Hence, the object could be one
of the about 10 missing globular clusters in the inner part of the Galaxy
(Ivanov et al. \cite{2005A&A...442..195I}). The circular optical appearance of
the cluster further supports this interpretation. However, to accurately
determine the cluster position in the Galaxy and the other parameters we need to
measure the age. This is not possible with the currently available dataset and
requires much deeper observation down to the cluster main sequence.

\begin{table}
\centering
\caption{\label{properties} Measured properties of the cluster FSR\,1735
(assuming a 12\,Gyr population). Listed
are the right ascension, declination, Galactic longitude, Galactic latitude,
K-band extinction, metallicity, core radius, tidal radius, central star density,
distance to the Sun, distance to the Galactic centre, distance to the 
Galactic plane, diameter, absolute K-band magnitude and the cluster mass.}
\begin{tabular}{lr}
Parameter & Value 		\\
\hline
R.A. (J2000) & 16 52 10.6 	\\
DEC (J2000) & $-$47 03 29 	\\
$l$ [deg] & 339.1879 		\\
$b$ [deg] & $-1.8534$ 		\\
$A_K$ [mag] & 0.5$\pm$0.1 	\\
$[$M/H$]$ & $-0.8\pm$0.1 	\\
R$_{core}$ ["] & 23.5		\\
R$_{tid}$ ["] & 86		\\
I$_{cen}$ [**/arcsec$^2$] & 0.27\\
R$_\odot$ [kpc] & 9.1$\pm$1.0	\\
R$_{GC}$ [kpc] & 3.2$\pm$0.6	\\
z$_{GP}$ [pc] & 290		\\
d [pc] & 2.1			\\
$M_K$ [mag] & $-8.2\pm$0.2	\\
M$_{cluster}$ [M$_\odot$] & 65000\\
\end{tabular}
\end{table}

\section{Conclusions}

We present the new Milky Way globular cluster candidate FSR\,1735 from a
2MASS-based near infrared cluster search (Froebrich  et al. 
\cite{2006MNRAS.subm.F}). The analysis is based on deep JHK images  taken with
Sofi at the NTT. The images show a rich circular cluster of stars. Radial star
density profiles show that the more luminous stars ($R_{core}$=23.5") are more
concentrated towards the cluster centre than the fainter stars
($R_{core}$=41.5"). This can be explained by magnitude migration due to
crowding. The reddening to the cluster is estimated to $A_K$=0.5$\pm$0.1\,mag,
mostly caused by a single cloud of gas and dust in the line of sight. From (J-K)
vs. K colour magnitude diagrams  we determine a metallicity of
[M/H]=$-0.8\pm$0.1. 

The significant peak in the K-band luminosity function of the cluster is
interpreted as the clump of horizontal branch stars. Based on this we estimate a
distance for the object of 9.1$\pm$1.0\,kpc, an absolute brightness of
$M_K=-8.2\pm$0.2\,mag and a diameter of 2.1\,pc. The number of horizontal branch
stars is used to estimate a cluster mass of about 6.5$\cdot$10$^4$\,M$_\odot$.

The lack of indications of on-going or recent star formation in the cluster (no
YSOs, IRAS or MSX sources, radio continuum sources) lets us to conclude that the
object is not a young open cluster. The position in the Galaxy, i.e. the
Galactocentric distance, poses strong arguments against the interpretation of
FSR\,1735 as an old open cluster. Only an at maximum 0.5\,Gyr old open cluster
with [M/H]=$-0.8$ could explain the observations, which would be, however, 
clearly at variance with the age-metallicity relation for the Galactic disc.

All the available observational evidence is in agreement with the interpretation
that FSR\,1735 is a so far unknown globular cluster in the inner Milky Way. 
A definite determination of the object parameters, however, requires to
measure the age of the cluster accurately, which is not possible with the
currently available dataset.

\section*{acknowledgments}

We would like to thank the referee Sergio\,Ortolani for helpful comments.

\label{lastpage}

\end{document}